\documentclass[preprint2]{aastex}

\shorttitle{Dynamical and photometric imprints of feedback
processes on the early evolution of E/S0 galaxies.}
\shortauthors{Cirasuolo et al.}

\begin{document}
\def\simlt{\mathrel{\rlap{\lower 3pt\hbox{$\sim$}}\raise 2.0pt\hbox{$<$}}}
\def\simgt{\mathrel{\rlap{\lower 3pt\hbox{$\sim$}} \raise
2.0pt\hbox{$>$}}}
\def\lsim{\,\lower2truept\hbox{${<\atop\hbox{\raise4truept\hbox{$\sim$}}}$}\,}
\def\gsim{\,\lower2truept\hbox{${> \atop\hbox{\raise4truept\hbox{$\sim$}}}$}\,}
\def\be{\begin{equation}}
\def\ee{\end{equation}}
\def\Re{$R_e\:$}
\def\Me{$M_e(<R_e)\:$}
\def\Mv{$M_{\rm vir}\:$}
\def\Ms{$M_{\rm sph}\:$}
\def\Mbh{$M_{\rm BH}\:$}
\def\Msun{$M_{\odot}\:$}
\def\sv{${\rm \sigma/V_{\rm vir}}\:$}
\def\vv{$V_{\rm vir}\:$}
\def\zv{$z_{\rm vir}\:$}
\def\FJ{Faber-Jackson }

\title{Dynamical and photometric imprints of feedback processes on
the formation and evolution of E/S0 galaxies.}

\author{Michele Cirasuolo\altaffilmark{1}}
\email{ciras@sissa.it}
\author{Francesco Shankar\altaffilmark{1}}
\email{shankar@sissa.it}
\author{Gian Luigi Granato\altaffilmark{2,1}}
\email{granato@pd.astro.it}
\author{Gianfranco De Zotti\altaffilmark{2,1}}
\email{dezotti@pd.astro.it}
\and
\author{Luigi Danese\altaffilmark{1}}
\email{danese@sissa.it}

\altaffiltext{1}{International School for Advanced Studies,
SISSA/ISAS, Via Beirut 2-4, I-34014 Trieste, Italy}

\altaffiltext{2}{INAF - Osservatorio Astronomico di Padova, Vicolo
Osservatorio, I-35100 Padova, Italy}

\begin{abstract}
We show that the observed Velocity Dispersion Function of E/S0
galaxies matches strikingly well the distribution function of
virial velocities of massive halos virializing at $z\ge 1.5$, as
predicted by the standard hierarchical clustering scenario in a
$\Lambda$CDM cosmology, for a constant ratio $\sigma/V_{\rm
vir}\simeq 0.55\pm 0.05$, close to the value expected at
virialization if it typically occurred at $z\gsim 3$. This
strongly suggests that dissipative processes and later merging
events had little impact on the matter density profile. Adopting
the above $\sigma/V_{\rm vir}$ ratio, the observed relationships
between photometric and dynamical properties which define the
fundamental plane of elliptical galaxies, such as the
luminosity-$\sigma$ (Faber-Jackson) and the luminosity-effective
radius relations, as well as the $M_{\rm BH}$-$\sigma$ relation,
are nicely reproduced. Their shapes turn out to be determined by
the mutual feedback of star-formation (and supernova explosions)
and nuclear activity, along the lines discussed by Granato et al.
(2004). To our knowledge, this is the first semi-analytic
model for which simultaneous fits of the fundamental plane
relations and of the epoch-dependent luminosity function of
spheroidal galaxies have been presented.
\end{abstract}


%%%%%%%%%%%%%%%%%%%%%%%%%%%%%%%%%%%%%%
\section{INTRODUCTION}\label{intro}
%%%%%%%%%%%%%%%%%%%%%%%%%%%%%%%%%%%%%%
A major challenge for the present day astrophysical research is to
trace the evolution of the Universe from the formation of the
first stars in the so called ``dark ages'' to the present epoch,
explaining the variety of structures from galaxies to clusters.
Understanding the formation and evolution of spheroidal galaxies,
which comprise the most massive and most luminous galaxies in the
universe and contain a large fraction of all stars, turned out to
be particularly intriguing. Hierarchical models tend to predict
substantially fewer massive galaxies at high redshift than are
observed (Blain et al. 2002; Scott et al. 2002; Daddi et al. 2003;
Tecza et al. 2004; Somerville et al. 2004). On the other hand, the
traditional monolithic models, whereby these objects formed their
stars on a timescale shorter than their free-fall time and evolved
passively thereafter (Eggen, Lynden-Bell \& Sandage 1962), do not
fit into a coherent scenario for structure formation from
primordial density perturbations, and also tend to over-predict
high-redshift galaxies.

In this situation it is essential to look for guidance from
observational data. It has long been known that stellar
populations in elliptical galaxies are old and essentially coeval
(Sandage \& Visvanathan 1978; Bernardi et al 1998; Trager et al.
2000; Terlevich \& Forbes 2002). A color-magnitude relation is
also well established: brighter spheroids are redder (Bower et al.
1992). The widely accepted interpretation is that brighter objects
are richer in metals and the spread of their star formation epochs
is small enough to avoid smearing of their colors. The slope of
this relation does not change with redshift (Ellis et al. 1997;
Kodama et al. 1998) supporting this interpretation. The star
formation history of spheroidal galaxies is mirrored in the
Fundamental Plane (Djorgovski \& Davies 1987; Dressler et al.
1987) and in its evolution with redshift.  Elliptical galaxies
adhere to this plane with a surprisingly low orthogonal scatter
($\sim 15$\%), as expected for a homogeneous family of galaxies.
Recent studies (e.g. Treu et al. 2002; van der Wel et al. 2004;
Holden et al. 2004, 2005) suggest that ellipticals, both in the
field and in clusters, follow this fundamental relation up to $z
\sim 1$, consistent with the hypothesis that massive spheroids are
old and quiescent.

Direct evidence that massive galaxies with $M \simgt 10^{11}
M_{\odot}$ were in place at $z \simgt 2$, is provided by recent
$K$-band surveys (Cimatti et al. 2002; Kashikawa et al. 2003;
Fontana et al. 2005). The space density of Extremely Red Objects
(EROs) at  $z \simgt 3$ is only a factor $\sim 5$--10 less than at
$z \sim 1$ (Tecza et al. 2004).

While the evolution of dark matter halos is controlled only by
gravity, and therefore the underlying physics is simple (even if
the evolutionary behaviour is complex), processes involving
baryons are intricate and may hold the key to reconcile theory
with observations. As shown by Granato et al. (2004) the
feedback from supernova (SN) explosions and from nuclear activity
can reverse the hierarchical scenario for baryons (see also
Granato 2001). In other words, the canonical hierarchical Cold
Dark Matter (CDM) scheme - small clumps collapse first - is
reversed for baryon collapse and the formation of luminous objects
({\it Anti-hierarchical Baryon Collapse} (ABC) scenario).

In this paper, after a short overview of the Granato et al. (2004)
model (Sect.~\ref{sec:gds04}), we investigate the imprints of
processes governing the formation and the early evolution of
spheroidal galaxies on their local dynamic and photometric
properties, such as the Velocity Dispersion Function (VDF,
Sect.~\ref{sec_VDF}), the luminosity-velocity dispersion relation
(Faber \& Jackson 1976, Sect.~\ref{sec_FJ}), the stellar
luminosity-effective radius relation (Bernardi et al. 2003,
Sect.~\ref{sec_Re}), and the BH mass-velocity dispersion relation
(Ferrarese \& Merritt 2000; Tremaine et al. 2002; Onken et al.
2004; Sect.~\ref{sec_BH}). The main conclusions are presented and
discussed in Sect.~\ref{discuss_ch5}.

Throughout this work we adopt a spatially flat cold dark matter
cosmology with cosmological constant, consistent with the
Wilkinson Microwave Anisot\-ro\-py Probe (WMAP) data (Bennett et
al. 2003): $\Omega_m=0.3$, $\Omega_b=0.047$, and
$\Omega_\Lambda=0.70$, $H_0=70{\,\rm km~s^{-1}~Mpc^{-1}}$,
$\sigma_8=0.84$, and an index $n=1.0$ for the power spectrum of
primordial density fluctuations.

\section{Overview of the Granato et al. (2004) model} \label{sec:gds04}

While referring to the Granato et al. (2004) paper for a full
account of the model assumptions and their physical justification,
we provide here, for the reader's convenience, a brief summary of
its main features.

The model follows with simple, physically grounded recipes and a
semi-analytic technique the evolution of the baryonic component of
proto-spheroidal galaxies within massive dark matter (DM) halos
forming at the rate predicted by the standard hierarchical
clustering scenario within a $\Lambda$CDM cosmology. The main
novelty with respect to other semi-analytic models is the central
role attributed to the mutual feedback between star formation and
growth of a super massive black hole (SMBH) in the galaxy center.

The idea that SN and  QSO feedback play an important role in the
evolution of spheroidal galaxies has been pointed out by several
authors (Dekel \& Silk 1986; White \& Frenk 1991; Haehnelt,
Natarajan \& Rees 1998; Silk \& Rees 1998; Fabian 1999). Granato
et al. (2004) worked out, for the first time, the symbiotic
evolution of the host galaxy and the central black-hole (BH),
including the feedback. In this model, the formation rate of
massive halos ($2.5\times 10^{11} M_\odot \simlt M_{\rm vir}
\simlt 2\times 10^{13} M_\odot$) is approximated by the positive
part of the time derivative of the halo mass function (Press \&
Schechter 1974, revised by Sheth \& Tormen 2002). The gas, heated
at virial temperature and moderately clumpy (clumping factor
$\simeq 20$), cools to form stars, especially in the innermost
regions where the density is the highest. The radiation drag due
to starlight acts on the cold gas, further decreasing its angular
momentum and causing an inflow into a reservoir around the central
BH, to be subsequently accreted into it, increasing its mass and
powering the nuclear activity. In turn, the feedbacks from SN
explosions and from the active nucleus regulate the star formation
rate and the gas inflow, and eventually unbind the residual gas,
thus halting both the star formation and the BH growth. In fact,
important parameters are the efficiency of SN energy transfer to
the cold gas ($\epsilon_{\rm SN}$) and the fraction of the QSO
luminosity in winds.

Further relevant differences with other semi-analytic models are
the allowance for a clumping factor, $C\simeq 20$, speeding
up the radiative cooling so that even in very massive halos
($M_{\rm vir}\sim 10^{13}\,M_\odot$), the gas, heated to the
virial temperature, can cool on a relatively short ($\sim
0.5$--$1\,$Gyr) timescale, at least in the dense central regions,
and the assumption that the large-scale angular momentum does not
effectively slow down the collapse of the gas and the star
formation in massive halos virialized at high redshift. As for the
latter assumption, we note that during the fast collapse phase,
when the potential well associated with a galactic halo is
established, gas clouds can lose their orbital energy to the dark
matter by dynamical friction on a timescale shorter than the halo
collapse timescale (see, e.g., Mo \& Mao 2004).

The model prescriptions are assumed to apply to DM halos
virializing at $z_{\rm vir} \gsim 1.5$ and $M_{\rm vir} \gsim 2.5
\times 10^{11} M_\odot$. These cuts are meant to crudely single
out galactic halos associated with spheroidal galaxies. Disk (and
irregular) galaxies are envisaged as associated primarily to halos
virializing at $z_{\rm vir} \lsim 1.5$, some of which have
incorporated most halos less massive than $2.5 \times
10^{11}\,M_\odot$ virializing at earlier times, that may become
the bulges of late type galaxies. However, the model does not
address the formation of disk (and irregular) galaxies and these
objects will not be considered in this paper.

The kinetic energy fed by supernovae is increasingly effective,
with decreasing halo mass, in slowing down (and eventually
halting) both the star formation and the gas accretion onto the
central black hole. On the contrary, star formation and black hole
growth proceed very effectively in the more massive halos, giving
rise to the bright SCUBA phase, until the energy injected by the
active nucleus in the surrounding interstellar gas unbinds it,
thus halting both the star formation and the black hole growth
(and establishing the observed relationship between black hole
mass and stellar velocity dispersion or halo mass). Not only the
black hole growth is faster in more massive halos, but also the
feedback of the active nucleus on the interstellar medium is
stronger, to the effect of sweeping out such medium earlier, thus
causing a shorter duration of the active star-formation phase.

The basic yields of the model are the star-formation rate,
$\psi(t)$, as a function of the galactic age, $t$, (hence the
evolution of the mass in stars, $M_{\rm sph}(t)$), and the growth
of the central BH mass, $M_{\rm BH}(t)$, for any given value of
the halo mass, $M_{\rm vir}$, and of the virialization redshift,
$z_{\rm vir}$. These quantities are obtained solving the system of
differential equations given by Granato et al. (2004), i.e. their
eqs. (10), (16), and (23). 
The rates at which the diffuse gas cools 
[$\dot{M}_{\rm cold}(t)$] and forms stars are given by their eqs. (9) and
(10), respectively, while their eq. (16)
yields the growth rate -- due to the radiation drag mechanism -- 
of the central BH reservoir [$\dot{M}_{\rm inflow}(t)$],
allowing the building-up of the BH mass (eq. 23).  
Finally, the rates at which the feedback from SN [$\dot{M}_{\rm cold}^{\rm SN}(t)$]
and QSO [$\dot{M}_{\rm inf}^{\rm QSO}$] heats the cold gas moving it into the 
hot phase are given by their eqs. (11) and (31), respectively.
In their notation:
\begin{eqnarray}
\dot{M}_{\rm inf}(t)&=&-\dot{M}_{\rm cold}(t)+\dot{M}_{\rm
inf}^{\rm QSO} \\
M_{\rm gas}(t)&=&M_{\rm inf}(t)+M_{\rm cold,sf}(t) \\
\dot{M}_{\rm res}(t)&=& \dot{M}_{\rm inflow}(t) -\dot{M}_{\rm BH}(t)  \\
\dot{M}_{\rm cold,sf}(t)&=&\dot{M}_{\rm
cold}(t)-\psi(t)  +\dot{M}_{\rm cold}^{\rm SN}(t) \nonumber \\
 & &+\dot{M}_{\rm
cold}^{\rm QSO}(t),
\end{eqnarray}
where a dot above a symbol denotes a time derivative. 
Note that, in eq.~(10) of Granato et al. (2004), defining
$\psi(t)$, ${M}_{\rm cold}(r,t)$ is actually ${M}_{\rm
cold,sf}(r,t)$.

Once the star-formation history, $\psi(t)$, of a galaxy has been
computed, its luminosity in any chosen band is obtained, as a
function of $t$, from the GRASIL code (Silva et al. 1998), which,
given $\psi(t)$, yields the chemical and the spectro-photometric
evolution from the radio to the X-ray band, allowing for the
effect of dust absorption and reradiation. GRASIL is available at
{\sl web.pd.astro.it/granato} and at {\sl
adlibitum.oat.ts.astro.it/silva/default.html}. As mentioned above,
the halo formation rate is obtained using the Press \& Schechter
(1974) formalism, as improved by Sheth \& Tormen (2002; see
eq.~(5) of Granato et al. 2004).

Putting these ingredients together we obtain the model predictions
for the statistical properties of spheroidal galaxies as a
function of cosmic time. Although the model has several
parameters (listed in Table 1 of Granato et al. 2004), all of them
are strongly constrained by the wealth of observational data that
have become available in recent years, including the luminosity
functions of spheroidal galaxies at different redshifts, their
counts and redshift distributions in optical and near-IR bands,
the $850\,\mu$m SCUBA counts and the associated (if preliminary)
redshift distributions, the relationship between the black-hole
mass and the stellar velocity dispersion, the local black-hole
mass function, and more (Granato et al. 2004; Silva et al. 2005;
Shankar et al. 2004).

\begin{figure}[t]
\epsscale{1} \plotone{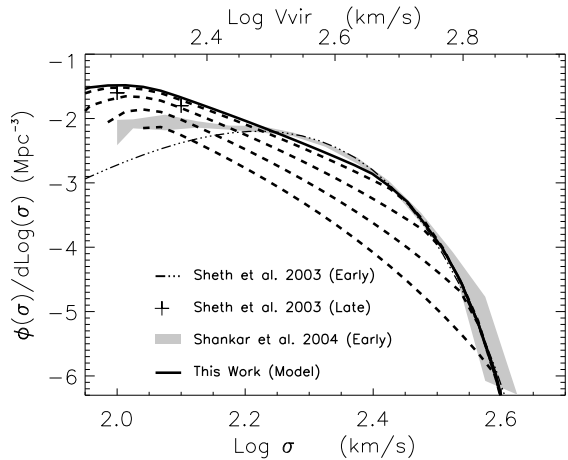} \caption{Comparison of the virial
velocity function of galaxies with $z_{\rm vir} \ge 1.5$ implied
by the standard gravitational clustering scenario (solid line,
upper scale), with observational estimates of the velocity
dispersion function of early-type galaxies by Shankar et al.
(2004; shaded area) and Sheth et al. (2003; triple-dot dashed
line). The crosses show the contribution of bulges of late-type
galaxies, determined by Sheth et al. (2003). The two functions
match for $\sigma =0.55 V_{\rm vir}$. The dashed lines show the
virial velocity functions of galaxies for different choices of the
minimum $z_{\rm vir}$: from top to bottom, $z_{\rm vir,min}=2$, 3,
4, 5. } \label{vdf}
\end{figure}

%%%%%%%%%%%%%%%%%%%%%%%%%%%%%%%%%%%%%%%%%%%%
\section{Properties of early type galaxies}
%%%%%%%%%%%%%%%%%%%%%%%%%%%%%%%%%%%%%%%%%%%%%

%%%%%%%%%%%%%%%%%%%%%%%%%%%%%%%%%%%%%%%%%%%%%%%%%%%%%%
\subsection{The Velocity Dispersion  Function}\label{sec_VDF}
%%%%%%%%%%%%%%%%%%%%%%%%%%%%%%%%%%%%%%%%%%%%%%%%%%%%%%

As discussed by Loeb \& Peebles (2003), the local Velocity
Dispersion Function (VDF) can provide interesting hints on the
structure formation process\footnote{Throughout this paper the
line-of-sight velocity dispersion $\sigma$ is referred to $R_e/8$,
where the effective radius $R_e$ is defined as the radius
containing half of the light.}. Accurate determinations of the VDF
of early-type galaxies have been obtained by Sheth et al. (2003)
and Shankar et al. (2004), based on a large sample ($\sim 9000$
E/S0 galaxies) drawn from the SDSS (Bernardi et al. 2003). The two
determinations  agree remarkably well (see Fig.~\ref{vdf}) except
for low values of the velocity dispersion, where the Sheth et al.
(2003) VDF declines due to the selection criteria adopted by
Bernardi et al. (2003) to compute the low luminosity portion of
the luminosity function.

As for dark matter halos, it is convenient to define a 'virial'
velocity, equal to the circular velocity at the virial radius
(Navarro et al. 1997, NFW; Bullock et al. 2001):
\begin{equation}\label{eq_vvir}
V_{\rm vir}^2=\frac{GM_{\rm vir}}{R_{\rm vir}}\ .
\end{equation}
Since, given the virialization redshift, $V_{\rm vir}$ depends
only on $M_{\rm vir}$ ($V_{\rm vir}\propto M_{\rm vir}^{1/3}$) the
$V_{\rm vir}$ distribution function can be straightforwardly
derived from the mass distribution function of spheroidal galaxies
[eq.~(5) of Granato et al. (2004)], integrated over the
virialization redshifts. Following Granato et al. (2004), we
assume that all massive halos ($2.5\times 10^{11} M_\odot \simlt
M_{\rm vir} \simlt 2\times 10^{13} M_\odot$) virializing at $z \ge
1.5$ yield spheroidal galaxies or bulges of later type galaxies.
It may be noted, in passing, that the adopted upper mass limit is
close to that inferred by Kochanek \& White (2001) from the
distribution of gravitational lens image separations.

As illustrated by Fig.~\ref{vdf}, the derived $V_{\rm vir}$
distribution function (which depends only on the evolution of dark
matter halos) accurately matches the observationally determined
VDF (which may be affected by the physics of baryons, and in
particular by dissipative processes) if $\sigma \simeq 0.55 V_{\rm
vir}$. The best fit and the confidence intervals of the
$\sigma/V_{\rm vir}$ ratio depend somewhat on the choice of the
upper mass limit, $M_{\rm sup}$ (reference value $M_{\rm sup}=
2\times 10^{13} M_\odot$) and of $\sigma_8$ (reference value
$\sigma_8=0.84$), while the choice of the minimum virialization
redshift ($z_{\rm vir,min}$) set at 1.5, does not affect
appreciably the fit. For $\sigma_8=0.84$, the 95\% confidence
interval, determined utilizing the $\chi^2$ statistic with 2
degrees of freedom, is $0.53\le \sigma/V_{\rm vir}\le 0.60$, the
upper end corresponding to the smallest value of $M_{\rm sup}$,
which is constrained to be $\ge 10^{13}\,M_\odot$ to ensure
consistency with the local $K$-band luminosity function (see
Granato et al. 2004). Decreasing $\sigma_8$ to 0.8, yields a 95\%
confidence interval $0.50\le \sigma/V_{\rm vir}\le 0.61$, while
increasing $\sigma_8$ to 0.9, we get an acceptable fit for $M_{\rm
sup} \simeq 10^{13}\,M_\odot$ or $\sigma/V_{\rm vir}\simeq 0.50$.
As already mentioned, the data indicate a $\sigma/V_{\rm vir}$
ratio independent of virial mass. A weak dependence, not steeper
than $\sigma/V_{\rm vir}\propto M_{\rm vir}^{0.05}$, is however
allowed.

A linear relationship between the central velocity dispersion,
$\sigma$, and the maximum circular velocity, $v_c^{\rm max}$
 -- which, for a concentration $c\simeq 3$ (see below), is
essentially equal to $V_{\rm vir}$ -- was reported by Gerhard et al.
(2001) for a sample of 21 mostly luminous, slowly rotating
elliptical galaxies, although the ratio is somewhat higher than
found here: $\sigma = 0.66 v_c^{\rm max}$. A weakly
non-linear relationship was found by Ferrarese (2002) for a sample
of 13 spiral galaxies with rotation curves extending beyond the
$B=25\, \hbox{mag}/\hbox{arcsec}^2$ isophote:
$\sigma/200\hbox{km}\,\hbox{s}^{-1} = 0.60
(v_c^{\rm max}/200\hbox{km}\,\hbox{s}^{-1})^{1.19}$. According to Ferrarese
(2002) this relation can be considered valid also in the
$\sigma$ range populated by elliptical galaxies.

If the stellar velocity dispersion profile is approximately
isothermal and stellar velocities are isotropic, adopting the
Navarro et al. (1997) density profile we obtain the following
relationship between $V_{\rm vir}$ and the velocity dispersion
$\sigma$:
\begin{equation}\label{eq_vsigma}
{\sigma\over V_{\rm
vir}}={\{c[3c^2+4c-2c\ln(1+c)-2\ln(1+c)]\}^{1/2}\over
\sqrt{6}[1+(1+c)\ln(1+c)]},
\end{equation}
where $c$, equal to the ratio of $R_{\rm vir}$ to the NFW inner
radius $r_s$, is the ``concentration''. The N-body simulations by
Zhao et al. (2003a) show that halos of mass greater than
$10^{11}\,\hbox{h}^{-1}\,M_\odot$ at $z\simgt 3$ have all a
similar median concentration $c\sim 3.5$.  For $c=2$, 3, 4,
eq.~(\ref{eq_vsigma}) yields $\sigma/V_{\rm vir} =0.49$, 0.57,
0.62, respectively. Thus, the value of $\sigma/V_{\rm vir}$ for
which we get a match between the local VDF and the $V_{\rm vir}$
distribution function of dark halos is remarkably close to the
value expected based on simulations.

The tight correspondence between the VDF and the velocity
distribution function of dark halos lends support to the dynamical
attractor hypothesis (Loeb \& Peebles 2003; Gao et al. 2004),
according to which the total distribution of collisionless matter
(dark matter plus stars) keeps essentially constant in the
presence of merging and of dissipative settling of baryons, with
the dark matter distribution expanding to compensate for the
dissipative settling of baryons (but see Gnedin et al. 2004 for a
different conclusion).

The stability of the stellar dynamics in the central regions of
dark halos against merging events subsequent to the virialization
redshift is also consistent with the results of detailed numerical
simulations (Wechsler et al. 2002; Zhao et al. 2003b) showing that
the halo circular velocity changes very little after the end of
the initial fast accretion process, during which most of the
specific binding energy is assembled, even though a large fraction
of the halo mass is acquired during the subsequent prolonged slow
accretion phase.

Figure 1 also shows the contributions to the VDF of different
virialization redshifts and highlights that the highest velocity
portion comes from the highest virialization redshifts. This
result is nicely consistent with the findings by Loeb \& Peebles
(2003) who computed the expected cumulative comoving VDF at $z=4$
for the NFW and the Moore et al. (1999) density profile, and found
it consistent with the observational determination by Sheth et al.
(2003) for velocity dispersions $\sigma >
300\,\hbox{km}\,\hbox{s}^{-1}$, while for lower values of $\sigma$
the predicted VDF is low compared to the observed one. Our curve
for $z_{\rm vir} \ge 4$ has a very similar behavior; but when we
allow also for spheroidal galaxies or bulges virializing at lower
redshifts we can fully account for the observed VDF. As mentioned
above, the contributions of sources virializing at $z\le 1.5$ is
negligibly small.

\begin{figure}[t]
\epsscale{1} \plotone{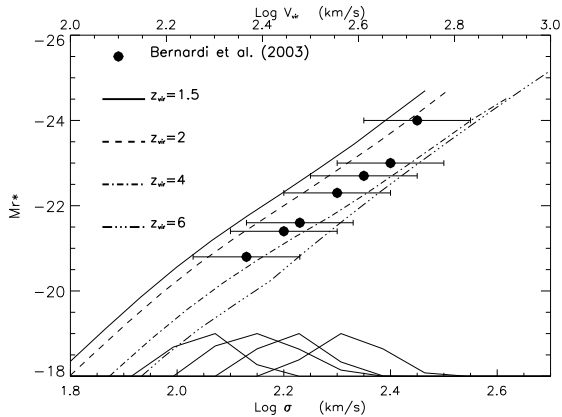}
\caption{Observed \FJ relation (filled dots) from
Bernardi et al. (2003), compared with  our predictions for
different virialization redshifts \zv. The curves at the bottom of
the figure represent the normalized distributions of velocity
dispersions of galaxies in 4 absolute magnitude bins 0.5 mag wide,
centered at $Mr*=-20.2$, $-21$, $-22$, and $-23$ (from left to
right), as predicted by the ABC model. The FWHMs of the predicted
distributions are remarkably close to the observed values
($\hbox{FWHM} \sim 0.09$; Bernardi et al. 2003).}
\label{FJ}
\end{figure}

%%%%%%%%%%%%%%%%%%%%%%%%%%%%%%%%%%%%%%%%%%%%
\subsection{The Faber - Jackson relation}\label{sec_FJ}
%%%%%%%%%%%%%%%%%%%%%%%%%%%%%%%%%%%%%%%%%%%%%%

Since the first measurements of velocity dispersions of early type
galaxies were made, it was recognized that they are correlated
with the galaxy luminosities (Poveda 1961; Minkowski 1962). Faber
\& Jackson (1976) showed that $L_{B}\propto \sigma^4$
(Faber-Jackson relation). Bernardi et al. (2003), using a sample
of $\sim 9000$ early type galaxies drawn from the Sloan Digital
Sky Survey (SDSS) in the redshift range $0.01 \le z \le 0.3$,
found $L_{r^*} \propto \sigma^{3.92}$, consistent with previous
studies (Forbes \& Ponman 1999; Pahre et al. 1998). Their data in
$u$, $g$, $i$, and $z$ bands show that the relation is roughly
independent of wavelength. The distributions of $\sigma$ at fixed
luminosity are approximately Gaussian.

A detailed quantitative analysis can be carried out using the ABC
model to follow the time evolution of the baryonic component, both
in the gas and in the stellar phase, of each halo mass $M_{\rm
vir}$ since the virialization redshift $z_{\rm vir}$. Then, the
present-day luminosity in any chosen band bands from X-ray to
radio can be computed with the spectrophotometric code GRASIL
(Silva et al. 1998) as a function of $V_{\rm vir}$, which is
uniquely determined by the value of $M_{\rm vir}$, given $z_{\rm
vir}$. As illustrated by Fig.~\ref{FJ}, the model not only
predicts the correct slope of the Faber-Jackson relation, but also
the correct normalization, when we use the ratio $\sigma/V_{\rm
vir}=0.55$ found from the analysis of the VDF. Acceptable fits can
be obtained with ${\rm 0.50 \simlt \sigma/V_{\rm vir} \simlt
0.65}$ (again, the confidence interval is derived using the
$\chi^2$ statistic).

The scatter in the observed relation is interpreted as an
intrinsic property of elliptical galaxies, accounted for by
different virialization epochs: galaxies with the same spheroidal
luminosity, but virializing at lower redshifts, have lower
velocity dispersions. The curves at the bottom of Fig.~\ref{FJ}
show the $\sigma$ distributions (arbitrary units) of galaxies in
four luminosity bins, as obtained from the Granato et al. (2004)
model. Such distributions have been computed integrating over
cosmic time, for each value of $\sigma$, the formation rate of
halos with present day luminosities within the considered bin;
they turn out to be roughly Gaussian, with the peak close to the
$\sigma$ expected from the best fit relation and FWHM in agreement
with that observed by Bernardi et al. (2003).

%The above result arises from the combination of the following
%circumstances.
It is worth noting that the standard scaling of the virial
parameters in the hierarchical clustering scenario gives (Bullock
et al. 2001) $M_{\rm vir}\propto V_{\rm vir}^3 (1+z_{\rm
vir})^{-3/2}$, which, for a roughly constant $M_{\rm vir}/L$
ratio, would imply a flatter slope than is observed in the \FJ
relation. However, the slope is steepened in the ABC model which
predicts a decrease of the $M_{\rm vir}/ M_{\rm sph}$ ratio
($M_{\rm sph}$ being the mass in stars), with increasing $M_{\rm
vir}$, whose details depend on the virialization redshift  (see
Fig.~5 of Granato et al. 2004). This is due to feedback from
supernovae, which is increasingly efficient with decreasing
$M_{\rm vir}$ in preventing the gas from cooling and forming
stars, tempered by the feedback from active nuclei which is more
effective in the more massive objects. On average, we have, to a
sufficient approximation, $M_{\rm vir}/M_{\rm sph} \propto M_{\rm
vir}^{-1/5}$. The ABC model also predicts an essentially constant
$M_{\rm sph}/L$ ratio (the observed weak luminosity dependence is
attributed to the systematic changes with luminosity of the
galactic structure, see Sect.~\ref{sec_Re}). Therefore, the
$M_{\rm vir}$--$V_{\rm vir}$ relation, for $\sigma/V_{\rm
vir}=\hbox{const}$,  translates into $L \propto \sigma^{18/5}
(1+z_{\rm vir})^{-9/4}$. The mean slope steepens somewhat,
approaching that of the observed Faber-Jackson relation, when we
account for the varying contributions of different virialization
redshifts to different $\sigma$ intervals. A further steepening of
the Faber-Jackson relation is expected at low $\sigma$ values,
corresponding to less massive objects where the SN feedback,
yielding $M_{\rm vir}/M_{\rm sph} \propto M_{\rm vir}^{-1/2}$,
dominates.

In conclusion, the Faber-Jackson relation is interpreted as
providing a quantitative measure of the effect of feedback, and
primarily of the effect of the energy injected onto the
interstellar medium by supernovae. The close agreement with the
predictions of the ABC confirms the correctness of the adopted
recipes. We note that fitting simultaneously the galaxy epoch
dependent luminosity functions and dynamical properties has long
been a very challenging problem for semi-analytic models (see,
e.g., Mo \& Mao 2004). In the case of spheroidal galaxies, this
requires that the model correctly predicts not only the evolution
of $L/M_{\rm vir}$ but also the distribution of virialization
redshifts as a function of the present-day luminosity. To our
knowledge, this is the first semi-analytic model for which
successful fits of all these quantities have been reported.

\begin{figure}[t]
\epsscale{1} \plotone{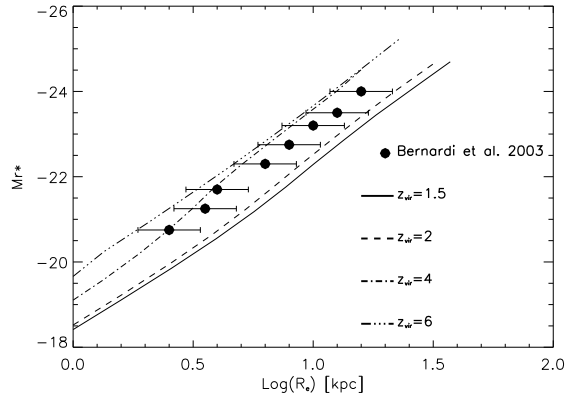}
\caption{Absolute magnitude in the $r^*$ band versus
$R_e$. The data points are from Bernardi et al. (2003) while the
lines show our model predictions for different virialization
redshifts.}
\label{lre}
\end{figure}

%%%%%%%%%%%%%%%%%%%%%%%%%%%%%%%%%%%%%%%%%%%%%%%%%%%
\subsection{Completing the Fundamental Plane}\label{sec_Re}
%%%%%%%%%%%%%%%%%%%%%%%%%%%%%%%%%%%%%%%%%%%%%%%%%%%
An additional, though not independent, check on the physical
processes involving baryons in the ABC model is provided by the
observed $L$--$R_e$ relation (see e.g. Bernardi et al. 2003).  The
effective radius, like the velocity dispersion, is related to the
collapse and settling of the baryonic component inside the DM
potential well. On the other hand, the model does not give us
dynamical information. However, it yields the total mass in stars
and dark remnants, while observations ensure that the starlight
distribution in spheroids has quite a uniform shape. In fact the
surface brightness distribution is well represented by
${\log(I(r))\propto r^{1/n}}$ (Sersic 1968); the classical de
Vaucouleurs profile has $n=4$ (de Vaucouleurs 1948).

Borriello et al. (2003), properly taking into account the light
distribution, the mass traced by light, and the dark matter, found
that the effective radius can be written as:
\be\label{re} R_e = (k_{\sigma}+\alpha_{\rm DM}) \frac{ G M_{\rm
sph}}{\sigma ^2},\ee
where $M_{\rm sph}$ is the total mass traced by light,
$k_{\sigma}$ is a constant depending on the light distribution
($k_{\sigma}$=0.174 for the de Vaucouleurs profile)  and
$\alpha_{\rm DM}$ is a function of the DM mass inside $R_e$ and of
its density distribution [see eqs. (13) and (14) of Borriello et
al. 2003]. As found by Gerhard et al. (2001) for a sample of about
20 elliptical galaxies and generalized by Borriello et al. (2003)
on the basis of the narrowness of the Fundamental Plane, the DM
inside the effective radius amounts to only 10-50\% of the total
mass. Such an amount and its distribution make $\alpha_{\rm DM}$
almost negligible in the above equation, that can be rearranged as
follows, using the definition of \vv [eq.~(\ref{eq_vvir})]:
\be\label{borri} \frac {R_e}{R_{\rm vir}} =
(k_{\sigma}+\alpha_{\rm DM}) \frac{ M_{\rm sph}/M_{\rm
vir}}{(\sigma/V_{\rm vir}) ^2},\ee
The ABC model yields the $M_{\rm sph}/M_{\rm vir}$ ratio and the
ensuing $M_{\rm sph}/L$ ratio, whose value depends  on the adopted
IMF. Using the Salpeter's IMF we get $M_{\rm sph}/L_r\simeq 5\,
M_\odot/L_\odot$, with very little dependence on the luminosity in
the reference $r^*$ band.

On the other hand, systematic changes with luminosity of the
galactic structure, quantified by the Sersic index $n$, have been
reported and found to account for the variation of the $M/L$ ratio
with luminosity (Ciotti et al 1996; Graham et al. 2001; Trujillo
et al. 2004). Bernardi et al. (2003), exploiting the large sample
of early-type galaxies in the SDSS, defined an effective mass
$M_0\equiv 4 R_e\sigma^2/G$, and found $M_0/L_{r}=3.5
({L_r}/{L_r^{\star}})^{0.15}$, in solar units, with
$L_r^{\star}=2\times 10^{10}L_\odot$. After eq.~(\ref{re})
$M_0=4(k_{\sigma}+\alpha_{\rm DM}) M_{\rm sph}$, so that,
neglecting $\alpha_{\rm DM}$ and setting $M_{\rm sph}/L_r\simeq
5\, M_\odot/L_\odot$, we have:
\be \label{ksigma} k_{\sigma} \simeq 0.174
(\frac{L_r}{L_r^{\star}})^{0.15}, \ee
consistent with the findings of Borriello et al. (2003) based on a
much smaller sample (221 nearby galaxies) but with more detailed
observations. Hence:
\be\label{borri2} \frac {R_e}{R_{\rm vir}} \simeq 0.87 \;
(\frac{L_r}{L_r^{\star}})^{0.15} \; \frac{ L_r/M_{\rm
vir}}{(\sigma/V_{\rm vir}) ^2}.\ee
The ABC model gives, for each value of $z_{\rm vir}$ and for an
assumed IMF, $R_{\rm vir}$ and $L_r$ as a function of $M_{\rm
vir}$. Using the above equation we can then obtain a relationship
between $L_r$ (or the absolute magnitude $M_{r^*}$) and $R_e$. The
results for a Salpeter IMF and $\sigma/V_{\rm vir}=0.55$, as
implied by the VDF, are compared in Fig.~\ref{lre} with the data
of Bernardi et al. (2003).  The presence of an average $\sim $30\%
of DM inside $R_e$ would imply $\alpha_{\rm DM} \leq 0.02$ (see
Borriello et al. 2003) and would not modify the fit.

The $M_{r^*}$--$R_e$ relation is related to the Faber-Jackson
relation through eq.~(\ref{re}), given the observationally
determined $L_r/M_{\rm sph}$ ratio. Thus, it does not provide an
independent test of the $L_{r}/M_{\rm vir}$ ratio predicted by the
ABC model. Still, the consistency of the model with a different
set of data is, at least, reassuring.

\begin{figure}[t]
\epsscale{1} \plotone{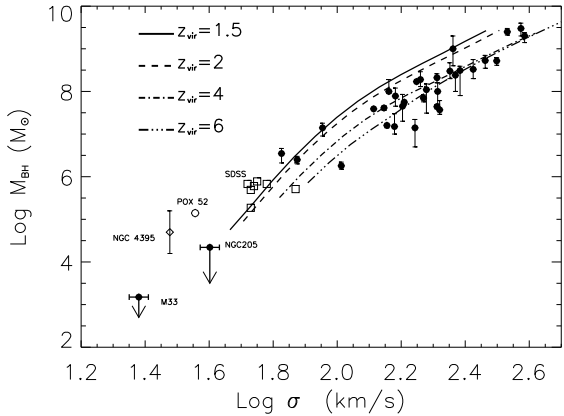} \caption{The $M_{\rm BH} - \sigma$
relation predicted by the model for different virialization
redshifts compared with observational data. Filled circles
represent nearby galaxies with dynamical measurements of the BH
mass (Tremaine et al. 2002) and the upper limits for M33 (Gebhardt
et al. 2001) and NGC~205 (Valluri et al. 2005). Open symbols refer
to galaxies with BH masses estimated from the H$\beta$
linewidth-luminosity-mass relation: NGC 4395 (open diamond;
Filippenko \& Ho 2003), POX 52 (open circle; Barth et al. 2004)
and a  sample of 7 faint active nuclei drawn from the SDSS, (open
squares; Greene et al. 2004). } \label{mbhs}
\end{figure}

%%%%%%%%%%%%%%%%%%%%%%%%%%%%%%%%%%%%
\subsection{The central black hole}\label{sec_BH}
%%%%%%%%%%%%%%%%%%%%%%%%%%%%%%%%%%%%%
In the last years massive BHs were proven to be ubiquitous in
massive E/S0 galaxies (see e.g. Magorrian et al. 1998). Tight
relations have been discovered between BH masses, $M_{\rm BH}$,
and stellar velocity dispersions (Ferrarese \& Merritt 2000;
Gebhardt et al. 2000; Tremaine et al. 2002; Onken et al. 2004),
masses of the spheroidal components (McLure \& Dunlop 2002; Dunlop
et al. 2003; Marconi \& Hunt 2003), and masses of the dark halos
(Ferrarese 2002); see Ferrarese \& Ford (2004) for a comprehensive
review. The $M_{\rm BH}$--$\sigma$ relationship has been
extensively discussed as a probe of the interplay between QSO and
host galaxy evolution (Silk \& Rees 1998; Fabian 1999; Cavaliere
et al. 2002).

\begin{figure}[t]
\epsscale{1} \plotone{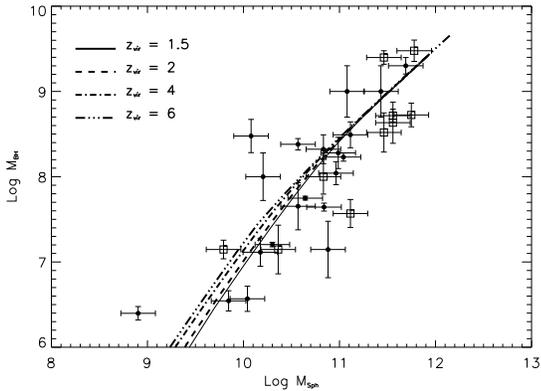}
\caption{ Predicted \Mbh-\Ms relation for different
virialization redshifts compared with observational data from
H\"{a}ring \& Rix (2004). Open squares are sources from their
Group 1 and solid circles from their Group 2.}
\label{nfp}
\end{figure}

Granato et al. (2004) showed that the growth of the stellar
component and of the central BH are strictly symbiotic, double
tied by the positive effect of the photon drag, proportional to
the star formation rate, favoring the inflow toward the reservoir
around the central BH (Umemura 2001; Kawakatu \& Umemura 2002),
and by the negative effect of the BH growth, powering outflows.

The ABC model follows, for any given \Mv and $z_{\rm vir}$, the
growth of the central BH and gives its final mass. The predicted
$M_{\rm BH}$--$V_{\rm vir}$ relation, at fixed $z_{\rm vir}$, is
immediately translated into $M_{\rm BH}$--$\sigma$ using the ratio
$\sigma/V_{\rm vir}=0.55$ derived from the VDF. The scatter of
data points is interpreted as reflecting the distribution of
$z_{\rm vir}$.

The model also predicts a steepening of the relation at low
$\sigma$ values, due to the combined effect of SN feedback --
which is increasingly efficient with decreasing halo mass in
slowing down the gas infall onto the central BH -- and of the
decreased radiation drag (due to a decrease of the optical depth).
From an observational point of view, the behavior of the $M_{\rm
BH}$--$\sigma$ relation in the low BH mass and low velocity
dispersion regime is still unclear, due to the dearth of data and
to the uncertainties on the \Mbh estimates. In Fig.~\ref{mbhs} we
plot objects found in literature with estimated BH masses $M_{\rm
BH} \simlt 10^6 M_{\odot}$. Dynamical measurements are available
only for M33 and NGC~205, while all the other  BH masses have been
estimated through the linewidth-luminosity-mass scaling relation
(Kaspi et al. 2000). Therefore the cases of M33 and NGC~205 are
particularly interesting, although they are outside the range of
masses to which we applied our model up to now. For M33 the upper
limit to the BH mass is $M_{\rm BH} \simlt 3000 M_{\odot}$
(Merritt et al. 2001) or $ M_{\rm BH}\simlt 1500 M_{\odot}$
(Gebhardt et al. 2001), while, for NGC~205, $M_{\rm BH} \simlt
2.2\times 10^4 M_{\odot}$. Both fall below the extrapolation of
the $M_{\rm BH}$--$\sigma$ relation holding at higher $\sigma$
(Ferrarese \& Merritt 2000; Gebhardt et al. 2000; Tremaine et al.
2002; Onken et al. 2004), while it is fully consistent with the
extrapolation of our model. On the other hand, the BH mass
estimates for the faint type 1 Seyfert nuclei in NGC 4395
(Filippenko \& Ho 2003), in POX 52 (Barth et al. 2004) and in 7
galaxies drawn from SDSS (Greene et al. 2004) are only marginally
consistent with the steepening predicted by our model. More data
on the so called intermediate mass BHs ($10^3 M_{\odot} \simlt
M_{\rm BH} \simlt 10^6 M_{\odot}$)  in the galactic centers (see
van der Marel (2003) for a review) are needed to clarify this
issue. Of course we should also keep in mind that the low BH mass
portion of the diagram might just reflect the distribution of BH
seeds (possibly created by merging of smaller BHs during the fast
accretion epoch) and be only weakly affected by the mass accretion
which is controlled by the effects discussed above. For example,
Koushiappas et al. (2004) presented a model yielding seed BHs with
characteristic masses $\sim 10^5\,M_\odot$. Also, low BH
masses may be easily increased by substantial factors during later
accretion phases, when host galaxies are disk- (rather than
bulge-) dominated.

Another prediction of the ABC model is that the ratio of the BH
mass to the mass in stars, $M_{\rm BH}/M_{\rm sph}$, is almost
insensitive to variations of $z_{\rm vir}$.  Therefore the $M_{\rm
BH}$--$M_{\rm sph}$ relation is expected to have a smaller {\it
intrinsic} scatter (see Fig.~\ref{nfp}) than the $M_{\rm
BH}$--$\sigma$ relation. This is because, as mentioned above, the
growth of the BH mass is controlled by the star formation rate
through the radiation drag and the SN feedback, and in turn, the
feedback from the active nucleus can eventually sweep out the gas
thus halting both the star formation and the accretion on the BH.
Thus the stellar and BH mass grow (and stop growing) in parallel.
The parallelism is not exact, however, since the star formation
rate has a twofold effect on the BH growth. As a result, the
$M_{\rm BH}$--$M_{\rm sph}$ relation is not strictly linear, but
bends down at small masses, and is slightly different for
different values of $z_{\rm vir}$. On the other hand, estimates of
$M_{\rm sph}$ are somewhat indirect and therefore liable to larger
uncertainties than those of $\sigma$, that can be directly
measured; this may translate in an {\it observed} scatter around
the mean $M_{\rm BH}$--$M_{\rm sph}$ relation comparable to, or
larger than that for the $M_{\rm BH}$--$\sigma$ relation, in spite
of the smaller intrinsic scatter.

In Fig.~\ref{nfp} we plot the \Mbh-\Ms relation for objects with
reliable bulge mass determinations from H\"{a}ring \& Rix (2004).
The agreement between the data, suggesting a slightly non-linear
relation ($M_{\rm BH} \propto M_{\rm sph}^{1.12}$), with our
predictions is remarkably good. As pointed out by H\"{a}ring \&
Rix (2004), a significant fraction of the scatter (which is
$\simlt 0.3$ dex) can be attributed to measurement errors. Marconi
\& Hunt (2003) found that the scatter in this relation is reduced
to $\sim 0.25$ dex, when \Ms is estimated as a virial mass ($\sim
R_e \sigma^2$).

The very good fits of the observed $M_{\rm BH}$--$\sigma$ and
\Mbh--\Ms relations are additional strong indications that the ABC
model properly includes the mutual feedbacks of stars and QSOs. It
is worth noticing that the model also correctly predicts the local
BH mass function in spheroidal galaxies (Granato et al. 2004;
Shankar et al. 2004).

\begin{figure*}[t]
\epsscale{1.9} \plotone{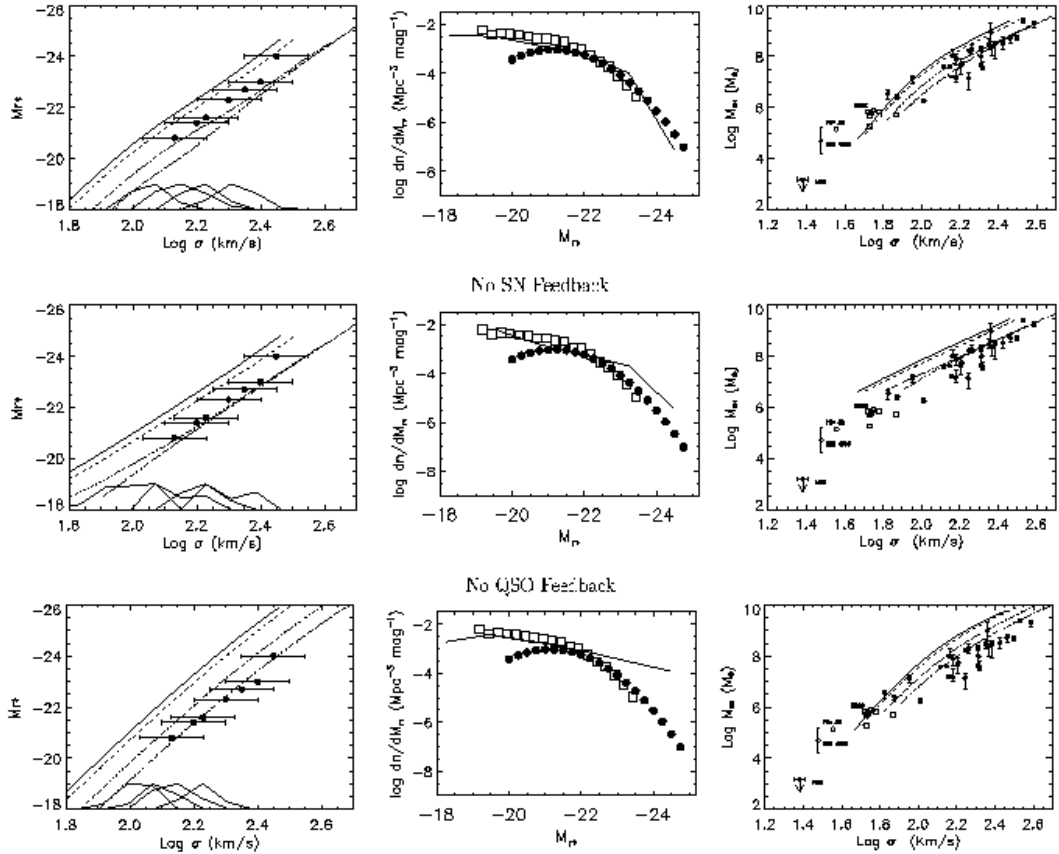}
%\epsscale{1.} \plotone{f6.eps}
\caption{ Effect of feedback on the Faber-Jackson relation (left
panel column), on the local $r^*$-band luminosity function of
spheroidal galaxies (central panel column), and on the $M_{\rm
BH}$--$\sigma$ relation (right panel column), according to the ABC
model. The upper, central, and lower panel rows compare with the
data the ABC model predictions with feedback fully included and
switching off, in turn, the SN and the QSO feedback, respectively.
Symbols are as in Figs. \ref{FJ} and \ref{mbhs}. The observational
determinations of the local luminosity function are from  Nakamura
et al. (2002; open squares), and Bernardi et al. (2003; filled
circles). } \label{fig:multi}
\end{figure*}

%%%%%%%%%%%%%%%%%%%%%%%%
\section{Discussion and conclusions}\label{discuss_ch5}
%%%%%%%%%%%%%%%%%%%%%%%%%

We have pointed out an impressively, and unexpectedly, tight
correspondence between the virial velocities, $V_{\rm vir}$,
controlled by the dynamics of dark halos, and the stellar velocity
dispersions, $\sigma$, that feel the effect of dissipative baryon
loading. A straightforward comparison of the virial velocity
distribution, implied by the standard hierarchical clustering
scenario in a $\Lambda$CDM cosmology, with the observed velocity
dispersion function of spheroidal galaxies, shows that the two
functions match remarkably well for a constant ratio
$\sigma/V_{\rm vir}\simeq 0.55$. For a NFW density profile, this
ratio corresponds to a concentration parameter $c\simeq 3$,
noticeably close to that found from N-body simulations (Zhao et
al. 2003a) to apply to halos of mass greater than
$10^{11}\,\hbox{h}^{-1}\,M_\odot$ virializing at $z\simgt 3$, with
no significant dependence on halo mass. Thus the matter density
profile at virialization appears to be essentially unaffected by
the subsequent events, including mergers and the dissipative
contraction of baryons, consistent with the dynamical attractor
hypothesis (Loeb \& Peebles 2003; Gao et al. 2004).

The $V_{\rm vir}$--$\sigma$ relation is a key ingredient to
connect theoretical predictions with observations. Using the above
determination of the $\sigma/V_{\rm vir}$ ratio, we have shown
that the observed relationships between photometric and dynamical
properties of spheroidal galaxies, defining the Fundamental Plane,
carry clear imprints of the feedback processes ruling the early
evolution of spheroidal galaxies. The steeper slope of the
luminosity-$\sigma$ (Faber-Jackson) relation compared to the
predicted $M_{\rm vir}$--$V_{\rm vir}$ relation is interpreted as
due primarily to the heating of the interstellar medium by
supernovae, which increasingly hampers star-formation in smaller
and smaller halos, and to winds driven by the active nucleus, that
eventually sweep out the residual interstellar gas in the most
massive nuclei.  A full analysis shows  that the treatment of
feedback adopted in the ABC model nicely accounts not only for the
slope, but also for the normalization of the Faber-Jackson
relation (see Fig.~\ref{FJ}). The model implies that the observed
scatter is mostly intrinsic and due to the spread of virialization
redshifts. It also predicts a moderate steepening of the relation
at low masses. To our knowledge, this is the first
semi-analytic model for which simultaneous fits of the epoch
dependent luminosity functions and of the Faber-Jackson relation
have been reported.

The effect of feedback also determines the slope of the
luminosity-effective radius relation. As discussed in
Sect.~\ref{sec_FJ}, the feedback yields, approximately, $M_{\rm
vir}\propto M_{\rm sph}^{5/6}$. Then, taking into account
eq.~(\ref{borri}), neglecting $\alpha_{\rm DM}$ and with $k_\sigma
\propto L_r^{0.15}$, we get, from $M_{\rm vir} \propto R_{\rm
vir}^3$ (for given $z_{\rm vir}$), $R_e \propto L_r^{0.15} R_{\rm
vir}^{8/5}$ and $L_r \propto R_e^2$. Weighting with the redshift
distributions appropriate for each value of $R_e$ (high-$z$
galaxies contribute more to low $R_e$'s), the mean $L_r$--$R_e$
relationship flattens towards the observed relation $L_r \propto
R_e^{1.58}$ (Bernardi et al. 2003). A further flattening is
predicted at low $R_e$ values, where the SN feedback dominates,
yielding, approximately, $M_{\rm vir}\propto M_{\rm sph}^{2/3}$,
whence $R_e \propto L_r^{0.15} R_{\rm vir}^{5/2}$ and $L_r \propto
R_e^{1.4}$.  Again, the model correctly reproduces not only the
slope but also the normalization of the relation (see
Fig.~\ref{lre}). As in the case of the Faber-Jackson relation, we
expect a scatter comparable to the observed one due to the
different virialization redshifts. Although the $L_r$--$R_e$
and the Faber-Jackson relation do not provide independent tests of
the model (see Sect.~3.3) it is reassuring that two different data
sets consistently confirm that it yields the correct virial
mass-to-light ratio.

The evolution of the stellar component is tied to that of the
central black hole, and therefore the $M_{\rm BH}$--$\sigma$
relation is also shaped by the effect of feedback, primarily from
the active nucleus itself in the more massive systems, and from
supernovae in smaller objects. Again, observations are well
reproduced (see Fig.~\ref{mbhs}), the dispersion around the
best-fit relation is expected to be largely due to the different
virialization redshifts, and a steepening of the $M_{\rm
BH}$--$\sigma$ is expected for low BH masses, if these are mostly
due to accretion interrelated with the bulge formation.
However, small BH masses may be substantially increased by later
accretion, occurring when the host galaxy is disk-dominated.
Moreover, according to some analyses, seed BH masses may be $\sim
10^5\,M_\odot$, and the low-$\sigma$ portion of the diagram is
testing more the distribution of seed masses than the accretion
history.

We have further pointed out that the $M_{\rm BH}$--$M_{\rm sph}$
relationship is essentially independent of $z_{\rm vir}$ (see
Fig.~\ref{nfp}); its {\it intrinsic} scatter should therefore be
minimum, although this does not necessarily translate in a low
{\it observed} scatter due to the large uncertainties on the
$M_{\rm sph}$ estimates (compared to those on $\sigma$).

A synoptic view of the effect of feedback on the
luminosity-$\sigma$ relation, on the local $r^*$-band luminosity
function, and on the $M_{\rm BH}$--$\sigma$ relation, based on the
ABC model, is provided by Fig.~\ref{fig:multi}.  If we switch off
the supernova feedback (central panel row), we get a larger
luminosity at fixed virial velocity $V_{\rm vir}$. The increase is
of $\simeq$ 1 mag. for low luminosity/mass galaxies
($M_{r^*}\simeq -20$), where the stellar feedback is dominating
over the QSO feedback, and of 0.5 mag. at high luminosities
($M_{r^*}\simeq -23$). Turning off the QSO feedback (bottom panel
row) goes in the same direction, but now the effect is larger for
larger galaxies. To fit the observed \FJ relation with no stellar
or QSO feedback, we have to shift by 0.1 dex the $\sigma /V_{\rm
vir}$ ratio, but this is inconsistent with the observed VDF.

As shown by the central panel column of Fig.~\ref{fig:multi}, the
shift to higher luminosities occurring when the feedback is
switched off affects only weakly the low luminosity portion of the
luminosity function, because of its flat slope. On the other hand,
the high luminosity tail is very sensitive to it, and particularly
to the feedback from the active nuclei.

On the whole, the local luminosity function of galaxy spheroids
and the observed correlations between their properties provide
clear evidence that the feedback both from supernovae and from
active nuclei plays a key role in the evolution of these sources,
and yield rather stringent constraints on the parameters
controlling the coupling of the energy injected into the
interstellar medium.

\noindent
\section{\bf Acknowledgments}

We are grateful to the referee for useful comments. Financial
support from the Italian MIUR and INAF is acknowledged.

\end{document}